\documentclass[10pt,conference,nocompress]{IEEEtran}
\pdfoutput=1

\usepackage{hyperref}
\hypersetup{hidelinks}
\usepackage{cite}
\usepackage{xspace}
\usepackage{url}
\usepackage{tcolorbox}
\usepackage{subcaption}
\usepackage{multirow}
\usepackage{enumitem}
\usepackage{listings}
\usepackage{algpseudocode}
\usepackage[ruled,linesnumbered,vlined]{algorithm2e}
\lstdefinestyle{mystyle}{                
    numbers=left
}
\usepackage{amsmath}
\usepackage{graphicx}
\usepackage{booktabs}
\IEEEoverridecommandlockouts

\newcommand{\revised}[1]{\textcolor{black}{#1}}
\newcommand{\sql}[1]{\textcolor{black}{#1}}
\newcommand{\lz}[1]{\textcolor{black}{#1}}

\newcommand{\tool}[0]{\textit{SpecGen}\xspace}

\title{\textit{SpecGen}: Automated Generation of Formal Program Specifications via Large Language Models}
\begin{document}

\author{
    \IEEEauthorblockN{Lezhi Ma\textsuperscript{1}, Shangqing Liu\textsuperscript{1}\IEEEauthorrefmark{2}, Yi Li\textsuperscript{2}, Xiaofei Xie\textsuperscript{3}, and Lei Bu\textsuperscript{1}\IEEEauthorrefmark{2}\thanks{\IEEEauthorrefmark{2} Corresponding author.}}
    \IEEEauthorblockA{\textsuperscript{1} State Key Laboratory for Novel Software Technology, Nanjing University, P.R. China\\
    \textsuperscript{2} 
    Nanyang Technological University, Singapore\\
    \textsuperscript{3} 
    Singapore Management University, Singapore\\}
    \IEEEauthorblockA{lezhima@hotmail.com, shangqingliu666@gmail.com, yi\_li@ntu.edu.sg, xfxie@smu.edu.sg, bulei@nju.edu.cn
    }
}

\makeatletter
\patchcmd{\@maketitle}
  {\addvspace{0.5\baselineskip}\egroup}
  {\addvspace{-1\baselineskip}\egroup}
  {}
  {}
\makeatother


\maketitle


\begin{abstract}
In the software development process, formal program specifications play a crucial role in various stages, including requirement analysis, software testing, and verification. However, manually crafting formal program specifications is rather difficult, making the job time-consuming and labor-intensive. Moreover, it is even more challenging to write specifications that correctly and comprehensively describe the semantics of complex programs. To reduce the burden on software developers, automated specification generation methods have emerged. However, existing methods usually rely on predefined templates or grammar, making them struggle to accurately describe the behavior and functionality of complex real-world programs.

To tackle this challenge, we introduce \tool, a novel technique for formal program specification generation based on Large Language Models (LLMs). Our key insight is to overcome the limitations of existing methods by leveraging the code comprehension capability of LLMs. The process of \tool consists of two phases. The first phase employs a conversational approach that guides the LLM in generating appropriate specifications for a given program, aiming to utilize the ability of LLM to generate high-quality specifications. The second phase, designed for where the LLM fails to generate correct specifications, applies four mutation operators to the model-generated specifications and \sql{selects verifiable specifications from the mutated ones through a novel heuristic selection strategy by assigning different weights of variants in an efficient manner}. \lz{We evaluate \tool on two datasets, including the SV-COMP Java category benchmark and a manually constructed dataset containing 120 programs. Experimental results demonstrate that \tool succeeds in generating verifiable specifications for 279 out of 385 programs,} outperforming the existing LLM-based approaches and conventional specification generation tools like Houdini and Daikon. Further investigations on the quality of generated specifications indicate that \tool can comprehensively articulate the behaviors of the input program.
\end{abstract}


\vspace{-2mm}
\section{Introduction}

Formal specifications play a central role in describing, understanding, and reasoning about program behaviors.
They capture the intended or actual program behaviors, in terms of formal languages, with precise semantics. Formal specifications may take various forms, such as procedure pre-/post-conditions, loop invariants, and assertions at specific program locations.
They are essential in a variety of software quality assurance tasks, including software testing~\cite{Mesbah2012InvariantBasedTesting,nguyen2023tc4mt}, model checking~\cite{cabodi2008strengthening,beyer2015boosting}, and program verification~\cite{flanagan2002escjava,rodriguez2004program}.



Yet, a practical challenge is the absence of documented formal specifications in most real-world software projects, since manually writing high-quality specifications is highly nontrivial.
To alleviate the burden on software developers, several tools have been introduced for generating program specifications automatically~\cite{flanagan2001houdini,nimmer2002automatic,ernst2007daikon}, including Houdini~\cite{flanagan2001houdini} and Daikon~\cite{ernst2007daikon}, two most representative ones for Java programs. 
However, these tools rely heavily on predefined templates or grammars during the specification generation process. \revised{As claimed by Molina et al.~\cite{molina2022fuzzing}, the fixed templates involved result in a limited range of specifications covered, usually yielding overly simplistic specifications that struggle to capture the complex behaviors and functionalities of real-world programs accurately. This phenomenon poses non-negligible limitations for these tools, consequently hindering their applications in the actual software development process~\cite{ding2019leveraging,rahman2014comparative,wei2011inferring}.}

To address this challenge, we introduce \tool, an automated technique for Java program specification generation based on the Large Language Models (LLMs). With the rise of LLMs, extensive research has attempted to apply them in software engineering and LLMs exhibit outstanding performance in various tasks~\cite{hou2023large,xia2023conversational,ahmed2022few,zeng2022extensive,ma2023scope}, where LLMs have demonstrated remarkable capabilities on code comprehension and summarization~\cite{yuan2023evaluating}. Inspired by this insight, we believe that LLMs can serve as a potent solution to overcome the limitations of existing program specification generation methods. The core idea of this work is to leverage LLMs to generate specifications that accurately capture the real behaviors of input programs, thus imbuing these specifications with richer semantics for further practical use. 

The workflow of \tool comes in two phases. In the first phase, \emph{conversation-driven specification generation}, we aim to query the output specifications by conducting a conversation with the LLM. To start the conversation, a prompt is constructed with several few-shot examples for the initial query. During the conversation process, we utilize the verification failure information from the specification verifier as the feedback prompt for the next round of the conversation. In this way, LLMs receive more cues, facilitating them to better generate accurate specifications. Nevertheless, despite the powerful code understanding and generation capabilities of large language models, they still struggle to handle complex programs effectively i.e., generating accurate specifications for complex programs. \sql{Through our repeated observation and testing of the model-generated results, we found that although the generated content is not highly accurate, it is already very close to the oracle,} which motivates us to design the second phase, \emph{mutation-based specification generation}. It focuses on generating accurate specifications where the LLM fails to provide verifiable results. Specifically, given a verification failure result by the LLM, \tool endeavors to combine four different kinds of mutation operators to modify it and obtain all potential variants. A selector adopting a heuristic selection strategy by assigning different weights of variants further repeatedly chooses a subset of these mutated variants deemed most likely to pass the verification until the results are successfully verified.


\lz{To evaluate \tool, we conduct experiments on two datasets. We first evaluate \tool on the benchmark for the Java category of SV-COMP~\cite{svcomp}. To further evaluate the performance of \tool on different kinds of programs, we constructed another dataset containing 120 Java programs with manually written ground-truth specifications. The selected programs are highly representative, encompassing different control-flow structures and various data structures to avoid any bias in our evaluation. We compared the performance of \tool on the dataset against multiple baselines. The results of our evaluation demonstrate that \tool significantly outperforms the baseline methods. \tool successfully generated verifiable specifications for 279 out of the total 385 programs, outweighing 247 for AutoSpec~\cite{wen2024enchanting}, the best-performing LLM-based approach, and 98 for Houdini, the best-performing non-LLM method.} An ablation study on mutations was also conducted, proving the effectiveness of all four types of mutation operators. Additionally, the results of evaluations on the heuristic selection strategy suggested that our strategy effectively improves the efficiency of \tool compared to the random selection strategy. Furthermore, a user study was conducted to evaluate the semantic quality of the generated specifications, illustrating the ability of \tool to accurately and comprehensively characterize program behaviors. The main contributions are summarized as follows:

\begin{itemize}[leftmargin=*,topsep=2pt]
    \item \sql{A novel approach for formal program specification generation and corresponding prototype tool~\cite{github2024specgen}, leveraging the Large Language Models to generate accurate and comprehensive specifications to describe program behaviors. Benefiting from the code comprehension ability of LLMs, our approach is capable of generating specifications with high quality, overcoming the limitations of existing methods in generating simplistic and basic specifications.}
    \item \lz{A mutation-based generation approach to enrich the diversity of the LLM output, consisting of a set of mutation operators and a novel heuristic selection strategy proposed to improve the efficiency of the verification that existing works fail to consider.}
    \item A dataset \revised{named \textit{SpecGenBench},} with hand-written specifications by experts, facilitating follow-up research. Other than the established benchmark SV-COMP, we collected programs on a more diverse spectrum for deeper insights.
    \item A comprehensive evaluation to evaluate our approach in all aspects. We compare \tool against Purely LLM-based approaches and representative non-LLM approaches. \tool succeeds in 279 out of the 385 programs, significantly outperforming the baseline approaches. 
\end{itemize}

\vspace{-1mm}
\section{Background and Motivation}
\subsection{Specification Generation and Verification}\label{sec:background_invariant}

Program specifications encompass precise statements that describe the \lz{intended or actual behaviors} of a particular program, either in its entirety or in distinct parts. \lz{In this work, we focus on generating specifications for the actual behaviors of input programs.} A large proportion of program specifications are expressed in formal languages, such as mathematical expressions to describe the constraints on the behaviors of a program. There are different kinds of specifications such as pre-conditions, which establish constraints on function parameters, ensuring proper execution of the function, post-conditions, which delineate the properties of a set of variables that persist after a function is executed, and loop invariants, which represent a specialized form of specification, detailing properties that consistently hold before executing the loop body. 
For different programming languages, the specifications may have different implementation forms. For example, in Java, the specifications can be expressed in Java Modeling Language (i.e., JML)~\cite{burdy2005JML} where \verb|requires| statements denote the pre-conditions of a function, \verb|ensures| statements represent the post-conditions of a function, and \verb|maintaining| statements specify the loop invariants.

\begin{figure*}
    \centering
    \includegraphics[width=0.99\textwidth]{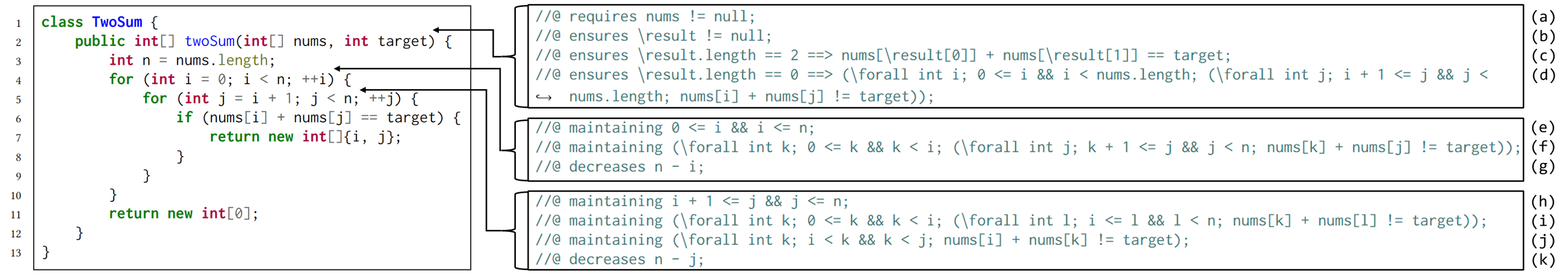}
    \caption{An example program and corresponding specifications generated by \tool, for which existing tools cannot generate comprehensive specifications to describe the program behaviors.}
    \label{fig:unique_example_program}
    \vspace{-4mm}
\end{figure*}

A series of automated program specification generation tools have been developed~\cite{flanagan2001houdini, nimmer2002automatic, ernst2007daikon, leonard2008automatic, pradel2009automatic} to reduce the burden on software developers. Two representative works are Houdini~\cite{flanagan2001houdini} and Daikon~\cite{ernst2007daikon}. Both rely on templates defined by human experts to generate a massive amount of candidate specifications, which verifiers then filter to eliminate incorrect candidate specifications until the remaining candidates are successfully verified. In particular, the template usually involves two or three variables and their corresponding operators i.e., \verb|<var> <op> <var>|, where \verb|<var>| should be filled in with variable names and \verb|<op>| should be an operator. For example, if a function contains two integer parameters \verb|x| and \verb|y| and an integer return value, Houdini may generate candidate pre-conditions and post-conditions for this function in the form of \verb|x > y|, \verb|x < y|, \verb|\result >= 0|, \verb|\result <= 0|, etc. where \verb|\result| is the defined variable denoting the return value in JML. For a program, all available variables within the scope such as class members, function parameters, and return values are taken to generate candidate specifications based on the defined templates and instrumented into corresponding points of the input program to verify the correctness of the specifications. The main difference between Houdini and Daikon lies in the design of the verifier where Houdini adopts a JML specification verifier\revised{, OpenJML~\cite{cok2011openjml},} which is designed from constraint solving~\cite{barrett2018satisfiability} for verifying. Yet, Daikon is based on the runtime checking which compares each candidate specification with the runtime execution traces.

\vspace{-1mm}
\subsection{\lz{Motivation}}\label{sec:motivation}

The existing automated program specification generation tools have limitations hindering real-world deployment and application. They rely on templates defined by human experts to generate specifications, which results in simple and trivial specifications. We present an example program on the left of Figure~\ref{fig:unique_example_program} for illustration. This program aims to search in the given integer array for the indexes of two separated elements of which the sum is exactly the given target value, and is implemented in two nested loops. If there do not exist such elements in the array, the program returns an empty array. To fully articulate the behaviors of the program, such properties must be specified, where Daikon and Houdini fail. In particular, both Houdini and Daikon can only generate trivial post-conditions such as \verb|nums != null| and \verb|\result[i] >= 0| for the method \verb|TwoSum()| as a whole. As for the outer-layer loop, only some simple loop invariants are generated, describing trivial numerical relationships between variables, such as \verb|i >= 0| and \verb|i < arr.length|. For the inner-layer loop, the generated specifications are \verb|j >= 1|, \verb|i < j| and \verb|j < nums.length|, which are similar to the out-layer. The generated specifications are too trivial, without detailed information to accurately capture the program's functionality.

Recently, large language models (i.e., LLMs)~\cite{GPT3.5, ChatGPT} have exhibited powerful capacities in coding~\cite{yuan2023evaluating, ma2023scope}. The emergence of these models may greatly compensate for the limitations of traditional software analysis tools in code understanding. A significant amount of work attempts to leverage large language models in software engineering~\cite{deng2023large, xia2023universal, jiang2023self, liu2023your, wei2023copiloting} and we have witnessed substantial progress brought about by the introduction of LLMs. Inspired by these works, in this work, we aim to leverage the large language models in the automated generation of formal program specifications to address the limitations of conventional \revised{template-based} approaches. From this perspective, we innovate our approach \tool, which generated three parts of specifications for the example presented in the right part of Figure~\ref{fig:unique_example_program}. The first part is to describe \revised{the} method \verb|TwoSum()| as a whole, specifying its pre-conditions and post-conditions. The specifications in lines a and b claim that the input array must not be null before and after the method is executed. The post-condition in line c specifies that the target value equals the sum of the two elements corresponding to the indexes stored in the returned array. The post-condition at line d specifies that there does not exist such a pair of elements that satisfies the constraint when the length of the returned array is zero. These generated specifications can fully articulate the functionality of method \verb|TwoSum()|. Furthermore, the loop invariants in the second and third parts specify corresponding constraints that must be met within a certain range in the array. These specifications generated by \tool comprehensively describe the semantics of this function and their correctness is verifiable.

\section{Approach}
\subsection{Overview}


\begin{figure*}[ht]
    \begin{minipage}[b]{0.6\linewidth}
        \centering
        \includegraphics[width=\textwidth]{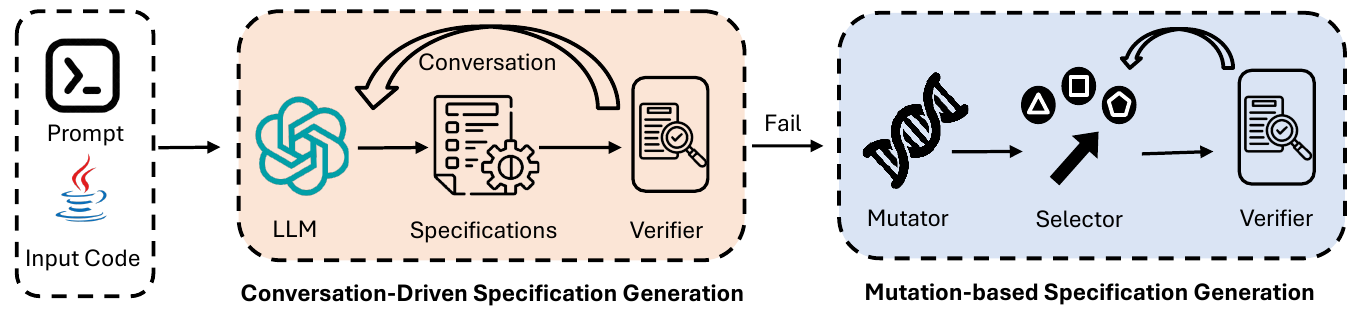}
        \caption{Overview of our \tool.}
        \label{fig:overview}
    \end{minipage}
    \begin{minipage}[b]{0.4\linewidth}
        \centering
        \includegraphics[width=0.93\textwidth]{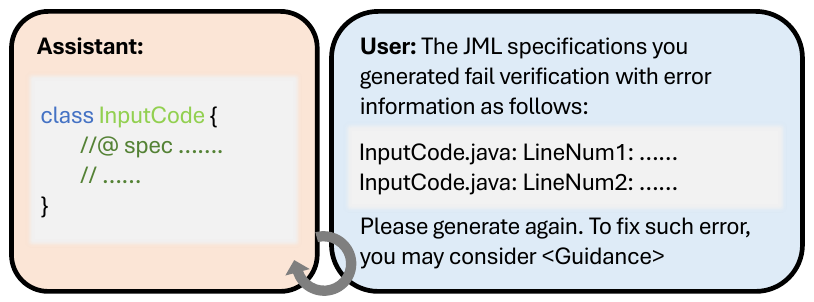}
        \caption{Conversational generation.}
        \label{fig:conversation-generation}
    \end{minipage}
    \vspace{-8mm}
\end{figure*}

The overview of \tool is presented in Figure~\ref{fig:overview}, which consists of two components i.e., conversation-driven specification generation and mutation-based specification generation. The former is designed to communicate with the large language model to query the output in a conversational manner. In particular, a prompt is constructed with some few-shot examples for the initial query. The verification failure information provided by the verifier is further used as the prompt for the next round of conversation if the model-generated results are incorrect. The conversation will be repeated iteratively until the generated specifications successfully pass the verifier or a maximum number of iterations is reached. The latter aims at generating the specifications of a program that the large language model fails to generate. Four kinds of mutation operators are adopted to mutate the specification that failed verification by the verifier and obtain all potential variants. A heuristic selector is further designed to efficiently choose a set of mutated variants most likely to pass verification.

\vspace{-1mm}
\subsection{Conversation-Driven Specification Generation}\label{sec:conversation-driven_generation}


\sql{Engaging in conversation with large models can fully leverage their capabilities, better assisting them in generating the desired content and avoiding potential errors~\cite{zhang2023siren}. Inspired by Xia et al.~\cite{xia2023conversational}, we propose our conversation-driven specification generation in \tool to interact with the LLM conversationally to generate specifications. There are two main benefits: firstly, the conversational manner aids the large model in automatically correcting potential syntax errors in the generated content, secondly, providing the model with the verification failures information by the verifier in the conversation helps it generate more accurate specifications.} The design mainly consists of two sequential components: initial prompt construction, which pre-defined an initial prompt to prepare for querying with the LLM, and conversational specification generation, which communicates to the LLM by incorporating verification failure information produced by the verifier in the conversation manner to generate verifiable specifications. The conversation will be repeated iteratively until the generated specifications pass the verifier or a maximum number of iterations is reached.

\begin{figure}[!t]
   \centering
   \includegraphics[width=0.47\textwidth]{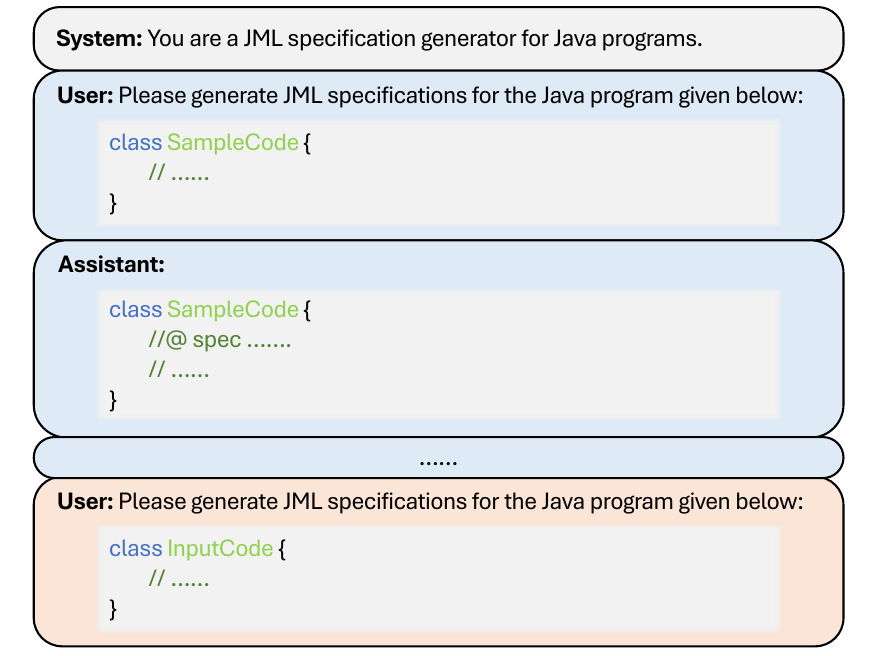}
   \caption{Illustration of the initial prompt construction.}
   \label{fig:prompt}
   \vspace{-6mm}
\end{figure}

\subsubsection{Initial Prompt Construction}
We need to define the initial prompt to query with LLM to obtain the model output. After multiple attempts to assess the impact of different prompts on the quality of generated program specifications, we ultimately chose to follow Xia et al.~\cite{xia2023keep} in designing our prompt. The prompt is presented in Fig.~\ref{fig:prompt} illustrating the components of the initial prompt, which consists of three different parts: the system role, few-shot examples, and the queried program. The system's role aims to inform LLMs about our application scenario, which is to generate JML specifications. We further add some few-shot examples. The reasons are two-fold. On one hand, few-shot examples can help the model to generate more accurate outputs~\cite{brown2020language}. On the other hand, LLMs can generate the desired output format that is learned from these examples. Each example is a pair of a program and its corresponding specifications. We randomly select it from our collected dataset to construct the few-shot examples. The last component is the queried program which requires the model to generate the output.


\subsubsection{Conversational Specification Generation}
Given the initial prompt, LLM can obtain the initial output for the input program. As the output of the LLM in the first attempt may not successfully pass the validation of the verifier, we interleave the process of specification generation with verification failure feedback to prompt future generation in a conversational manner which is illustrated in Fig.~\ref{fig:conversation-generation}. 
In particular, each generated specification by the model is verified by a JML verifier to test whether the generated result can pass the verifier. If the verification fails, we construct feedback information using the reported error message from the verifier as the prompt for the next generation. The verification error message can help the model understand the reason for failure and provide guidance for generating correct specifications. In addition, to avoid the verifier providing excessively long error messages, we configure the verifier to report only one verification failure message per attempt. Furthermore, through a massive amount of experiments, we summarize several types of common verification failures reported by the verifier. For each kind of error, we provide guidance in the natural language to facilitate the model in generating correct specifications. Upon encountering these types of verification failures reported by the verifier, we will insert corresponding guidance information into the prompt e.g., $<$Guidance$>$ in Fig.~\ref{fig:conversation-generation} to assist the model in resolving the issues. The conversation will be repeated iteratively until the specifications are successfully verified or a maximum number of iterations is reached.

\vspace{-1mm}
\subsection{\lz{Mutation-based Specification Generation}}\label{sec:palusible_invariant_generation}
In the conversation-driven generation process, \sql{some few-shot examples are provided in the initial prompt to start the query with the LLM. To further stimulate the potential of the LLM, multi-turn conversation continually guides the LLM in approaching the accurate specification more closely. Yet, they still struggle to generate fully correct specifications for some complex programs. The reasons are two-fold. On the one hand, in comparison to code generation~\cite{brown2020language, chen2021evaluating, openai2023gpt4}, specification generation poses greater challenges for LLMs due to the limited corpora related to program specifications for the model to learn from. On the other hand, although the verification failure information provided by the verifier can assist LLMs in providing higher-quality responses to some extent, as the error messages are highly abstract and generalized, LLMs still struggle to accurately understand the semantic information within error messages for complex programs. While LLMs may not accurately generate specifications for complex programs, the generated results are already highly close to the oracle, inspiring us to design mutation-based generation.}

In particular, the mutation-based specification generation component takes the output generated by the large language model that fails to pass the verifier through the multi-round conversation as the input. We further define a set of mutation operators to modify these generated outputs to obtain more diverse results. Then a heuristic strategy is adopted for efficient verification. The workflow is presented in Algorithm~\ref{alg:plausible_generation}. Specifically, we define the specifications generated by LLM that fail verification as the set of template specifications $E_t$, which consists of different specifications generated for different locations in a program, and a set of mutation operators as $M$. The \verb|MutationBasedGen| takes $E_t$ and $M$ as the input and outputs a set of correct specifications as $E$. The function \verb|SpecMutation| corresponds to the mutation operation of $E_t$, where each kind of mutation operator will be performed through the mutation function $Mutate()$ (Section~\ref{sec:invariant_mutation}) on a template specification $e$ ($e \in E_t$) to obtain a set of candidates $E_{mutated}$ (lines 7 and line 8). 

After the mutation operations are performed, we further design the specification selection algorithm to select a subset $E_{selected}$ of mutated specifications that can pass the verification. The selected subset $E_{selected}$ is initialized with $E_t$. We then iteratively require the verifier to check the correctness of $E_{selected}$ and obtain a set of refuted specifications denoted as $E_{refuted}$ from $E_{selected}$ that the verifier fails to verify. After that, we need to replace the failed specifications from $E_{refuted}$ with another mutated variant for the next iteration of verification. The $ReSelect$ function is presented from line 18 to line 25. For each refuted specification $e_r \in E_{refuted}$, we first remove it from the mutation set $E_{mutated}$ and the selected set $E_{selected}$. Then we replace $e_r$ with another mutated variant $e$ that comes from the same family by a heuristic selection strategy (Section~\ref{sec:spec-selection}) from line 22 to line 23. Here the family refers to a set of mutated specifications $E_f$ that come from the same template specification. Finally, we add $e$ to the selected set $E_{selected}$ to prepare for the next iteration of verification. The above process will be repeated until all candidates in $E_{selected}$ are successfully verified i.e., $E_{refuted}$ is empty (line 16). \lz{Note that if all candidates are refuted, the process will finally select an empty set of candidates, guaranteeing the termination of the whole process.}

\begin{algorithm}[!t]
    \scriptsize
    \SetKwFor{Loop}{loop}{}{EndLoop}
    \SetKwInOut{Input}{Input}
    \SetKwInOut{Output}{Output}
    \SetKw{Continue}{continue}
    \SetKw{Return}{return}
    \Input{Set of template specification $E_t$, set of mutations $M$}
    \Output{Set of verified specifications $E$}
    \SetKwFunction{PlausibleGen}{MutationBasedGen}
    \SetKwFunction{SpecMutation}{SpecMutation}
    \SetKwFunction{SpecSelection}{SpecSelection}
    \SetKwFunction{SelectNewSubset}{ReSelect}
    \SetKwProg{Myfunc}{Function}{}{}
    \Myfunc{\PlausibleGen{$E_t, M$}}{
        $E_{mutated} = \mathrm{SpecMutation}(E_t, M)$\\
        $E = \mathrm{SpecSelection}(E_{mutated}, E_t, M)$\\
        \Return $E$\\
    }
    \Myfunc{\SpecMutation{$E_t, M$}}{
        $E_{mutated} = \emptyset$\\
        \For{${e}\in {E_t}$}{
            $E_{mutated} = E_{mutated} \cup \mathrm{Mutate}(e,M)$\\
        }
        \Return $E_{mutated}$\\
    }
    \Myfunc{\SpecSelection{$E_{mutated}, E_t, M$}}{
        $E_{selected} = E_t$\\
        $E_{refuted} = \emptyset$\\
        \Repeat{$E_{refuted}$ is $\emptyset$}{
            $E_{refuted} = \mathrm{Verify}(E_{selected})$\\
            $E_{selected} = \mathrm{ReSelect}(E_{selected}, E_{mutated}, E_{refuted}, M)$\\
        }
        \Return $E_{selected}$\\
    }
    \Myfunc{\SelectNewSubset{$E_{selected}, E_{mutated}, E_{refuted}, M$}}{
        \For{$e_r \in E_{refuted}$}{
            $E_{mutated} = E_{mutated} \backslash \{e_r\}$\\
            $E_{selected} = E_{selected} \backslash \{e_r\}$\\
            $E_{f} = \mathrm{GetFamilyOf}(e_r, E_{mutated})$\\
            $e = \mathrm{SelectByHeuristic}(E_f, M)$\\
            $E_{selected} = E_{selected} \cup \{e\}$\\
        }
        \Return $E_{selected}$\\
    }
    \caption{Mutation-based Specification Generation}
    \label{alg:plausible_generation}
\end{algorithm}

\subsubsection{Template Specification Mutation}\label{sec:invariant_mutation}

As shown in Table~\ref{tab:mutations}, we define four kinds of mutation operators including predicative, logical, comparative, and arithmetic. Each type of mutation corresponds to one type of operator supported by JML. 
LLMs perform well in formulating the overall syntactical structure of specifications, but they often make mistakes in grasping the fine-grained relationships between variables, resulting in incorrect operators used to describe the relationships between variables, which is why our mutation design is centered around the operators. A mutation operation substitutes the operators of the corresponding type in the specification with another of the same type. For example, after applying a predicative mutation, a \verb|\exists| predicate within a specification may be substituted with \verb|\forall|.
Note that the mutation for a certain type of operator does not necessarily create only one mutated candidate. For example, the expression \verb|a <= b| may be mutated to \verb|a < b| or \verb|a - 1 <= b|. If multiple mutations can be applied to a specification at the same time, we try to exhaust each combination of different types of mutations to get all potential variants. \revised{Since the set of all potential variants of a certain template is determined, the exhaustive searching process is deterministic.} For instance, the expression \verb|x < n + 1| can be mutated to \verb|x <= n + 1| from the comparative type, \verb|x < n - 1| from the arithmetic type, or \verb|x <= n - 1| by combining them.


\begin{table}[!t]
\centering
\scriptsize
\vspace{-2mm}
\caption{The defined mutation operators.}
\label{tab:mutations}
\begin{tabular}{c|c|c}
\toprule
Mutation Type                & Original Operator               & Mutated Operators                     \\ \midrule
\multirow{2}{*}{Predicative} & \verb|\forall| & \verb|\exists|               \\ \cline{2-3} 
                             & \verb|\exists| & \verb|\forall|                \\ \hline
\multirow{5}{*}{Logical}     & \&\&                   & $\|$                                                    \\ \cline{2-3} 
                             & $\|$                     & \&\&
                             \\ \cline{2-3}
                             & \textless{}==\textgreater{}                   & \textless{}==, ==\textgreater{}                 \\ \cline{2-3}                             
                             & ==\textgreater{}                   & \textless{}== 
                             \\ \cline{2-3}                             
                             & \textless{}==                   & ==\textgreater{}
                             \\ \hline
\multirow{6}{*}{Comparative} & \textless{}=           & \textless{}, "- 1 \textless{}="       \\ \cline{2-3} 
                             & \textgreater{}=        & \textgreater{}, "+ 1 \textgreater{}=" \\ \cline{2-3} 
                             & \textless{}            & \textless{}=                          \\ \cline{2-3} 
                             & \textgreater{}         & \textgreater{}=
                                           \\ \cline{2-3} 
                             & ==            & !=
                                            \\ \cline{2-3} 
                             & !=         & ==                      
                                            \\ \hline
\multirow{2}{*}{Arithmetic}  & +                      & -                                     \\ \cline{2-3} 
                             & -                      & +                                     \\ \bottomrule
\end{tabular}
\vspace{-4mm}
\end{table}

\subsubsection{Mutated Specification Selection}\label{sec:spec-selection}
Typically, for a program, Houdini~\cite{flanagan2001houdini} verifies all generated specifications at one time. However, similar practice cannot be applied in \tool as we exhaust all potential combinations of mutations for a template specification. The set of the obtained specifications for verification is considerably large, posing a much greater burden for the verifier within a single verification process. To address this challenge, we innovate a heuristic selection strategy to improve the stability and efficiency of verification.




In general, the heuristic selection algorithm finds a specification $\hat{e}$ such that
\begin{equation}\label{eq:heuristic_selection}
    \hat{e} = \mathop{\arg\max}\limits_{e \in E_f}\sum\limits_{m \in M}(\mathop{times}(m,e,e_t) \cdot \mathop{weight}(m))
\end{equation}
where $E_f$ denotes a family of mutated specifications that come from the same template specification $e_t$, and $M$ denotes the set of all mutations. Given $E_f$, $e_t$, and $M$, we design the heuristic selection logic to prioritize selecting important candidates for verification. In particular, we assign scores for each mutated candidate $e \in E_f$ and select the candidate with the highest score as the output. To calculate the score of a candidate $e$, for all types of mutations $m \in M$, we sum up all the values of $times(m,e,e_t)$ multiplied by $\mathop{weight}(m)$, where $times(m,e,e_t)$ calculates how many times the mutation $m$ is performed when $e_t$ mutates into $e$, and $weight(m)$ denotes the corresponding weight of $m$.

\vspace{-1mm}
\section{Experimental Setup}
\vspace{-1mm}
We design the following four research questions for evaluation:
\begin{itemize}[leftmargin=*]
    \item \textbf{RQ1:} How does \tool compare with the baseline approaches?
    \item \textbf{RQ2:} How does each type of mutation contribute to the effectiveness of \tool?
    \item \textbf{RQ3:} How do different candidate selection strategies affect the efficiency of \tool?
    \item \textbf{RQ4:} To what extent can the generated specification contain the semantic information of the input program?
\end{itemize}

\vspace{-1mm}
\subsection{Implementation} \label{sec:implementation}
We use the API provided by OpenAI~\cite{GPTAPI} to communicate with the large language model of \verb|gpt-3.5-turbo-1106| for the experiments. Temperature is set to 0.4 to balance the diversity and rigorousness of the outputs of GPT. 4 few-shot examples are used during the prompt construction to balance the input length and response time. The maximum number of rounds of conversation is set to 10. The verifier is OpenJML~\cite{cok2011openjml}, the most recent JML specification verification tool to check the consistency between Java source code and JML specifications. 
Due to the incompleteness in the implementations of OpenJML, we set a timeout limit of 30 minutes for a single verification in our implementation to avoid unexpected situations, such as the non-responding of OpenJML. All experiments are conducted on an 8-core workstation with Intel Core i7-12700H CPU @2.30GHz and 32GB RAM, running Ubuntu 22.04.3 LTS. The version of OpenJDK is 1.8.0\_371 for all experiments except for Houdini, which has to run under OpenJDK 1.6.0\_45. We set the weight of comparative, logical, arithmetic, and predicative mutation to -1, -2, -4, and -4 respectively as the comparative mutation is more likely to pass the verification followed by the logical mutation. The predicative and arithmetic mutations are the least important through our observations from extensive experiments. Note that the weights are defined with negative values, leading to negative calculated scores as well. The reason for the design is to prioritize the specification candidates with fewer mutations.

\vspace{-1mm}
\subsection{Dataset}\label{sec:dataset}
\vspace{-1mm}

To comprehensively evaluate the effectiveness of \tool, following previous work~\cite{alshnakat2020constraint}, we first use an established dataset, the benchmark of SV-COMP~\cite{svcomp}, for evaluation. Specifically, we used 265 class definitions in the Java category of SV-COMP benchmark and conducted the necessary modifications on part of these programs (referred to as Dataset SV-COMP hereinafter) for ease of evaluation. The remaining data in the benchmark cannot be applied for specification generation even with our modification. \revised{We made minimal modifications to SV-COMP programs to ensure they can be executed outside the SV-COMP environment.} Specifically, the programs destined to trigger false assertions have to be modified so that the programs can exit properly. Also, those library calls specific to the competition settings (e.g. \verb|Verifier.nondetInt()|) are replaced with equivalent Java library calls so that they can be successfully compiled. \revised{It is ensured that the semantics of the modified programs remain unchanged.} \revised{As calculated by tool JaCoCo~\cite{eclEmma2024JaCoCo}, these programs have an average line of code (LoC) of 22.51, along with an average cyclomatic complexity (CC) of 6.18.} However, after a deep analysis of the characteristics of the data from SV-COMP, we find that 88.7\% programs are loop-free, indicating that the samples with more complex program structures cannot be covered by this dataset, inducing limitations on the evaluation. Also, very few datasets have been established specifically for specification generation tasks so far.

To remedy this gap, we further collect another dataset\revised{, \textit{SpecGenBench}, containing 120 samples as a supplement where 20 programs (including the corresponding specifications) from the dataset constructed by Nilizadeh et al.~\cite{Nilizadeh2021AprFormalMethods} and 100 programs from LeetCode~\cite{leetcode}.} The selected programs are assured of the feasibility of expressing their behaviors as JML-specified verifiable specifications. These programs involve a variety of control flow structures and encompass multiple data structures such as arrays, strings, and other data structures supported by the Java library. They also cover a diverse set of specifications including post-conditions and loop invariants, involving both linear and nonlinear relationships between variables, making them representative of a broad spectrum of scenarios. They can be categorized into five categories according to their types of control flow structures~\cite{xie2017automatic}. Specifically, \emph{Sequential} denotes the programs without branches or loops. \emph{Branched} represents the loop-free programs that will contain branches like if-else or switch structures. \emph{Single-path Loop} contains the simplest type of loop, with only one layer of loop structure without branches in their loop bodies. In contrast, \emph{Multi-path Loop} denotes the loops that have branches in the loop bodies. Lastly, \emph{Nested Loop} denotes the programs with multiple layers of loop structure where each layer may have a branch. The quantity of programs for \emph{Sequential}, \emph{Branched}, \emph{Single-path Loop}, \emph{Multi-path Loop} and \emph{Nested Loop} is 26, 23, 24, 26, and 21, respectively. \revised{Programs in \textit{SpecGenBench} have an average LoC of 20.77 and an average CC of 6.60.}

To obtain the ground truth specifications for the 100 Java programs from LeetCode, we follow a similar procedure in Nilizadeh et al.~\cite{Nilizadeh2021AprFormalMethods} with the help of human experts. Three experts with rich experience in formal verification were employed, to manually write specifications for each program. Each expert is required to write the specifications that can be successfully verified to describe the functionality and behavior of the program as accurately and comprehensively as possible. For a single program, if multiple experts have written verifiable specifications, another expert is responsible for selecting one of them as the ground truth.

As Daikon~\cite{ernst2007daikon} requires a set of test suites instrumented into the source code to execute the code, we manually write these test suites for it. \revised{A small test suite is initialized first, on which Daikon is invoked to generate specifications. If the results fail verification, the counterexample produced by the verifier will be added to the test suite. The procedure is repeated until no new specifications are generated. The test suites achieve an average instruction coverage of 90.16\%, branch coverage of 87.98\%, and line coverage of 91.36\%.} Lastly, we instrument dummy function calls at the top of each loop body in a program to ensure Houdini and Daikon can generate the specifications at these program points.

\vspace{-1mm}
\subsection{Baselines}\label{sec:baselines}
\revised{We select two conventional approaches and several LLM-based approaches as the baselines for comparison.}

\noindent \textbf{Houdini~\cite{flanagan2001houdini}}. It is a template-based JML annotation generator that relies on a series of pre-defined templates to generate candidate specifications. Given the input program, it first generates candidate specifications by filling in the templates with available variables and all kinds of operators. Afterward, It iteratively invokes a JML specification verifier to check their correctness and removes the refuted ones. The process will be repeated until the remaining candidates are all verified. 

\noindent \textbf{Daikon~\cite{ernst2007daikon}}. It is a classic tool for the dynamic detection of program specifications which relies on the dynamic execution trace of the target program to infer likely specifications. Given the input program, it first instruments the target program to trace certain variables and extracts execution traces. Then the inference engine reads the trace data and infers the potential invariants with a generate-and-check algorithm. Daikon supports dynamic detection for Java, C/C++, C\#, and Perl programs along with various formats such as DBC format~\cite{Jtest}, JML format~\cite{burdy2005JML}, and CSharpContract format~\cite{CodeContracts}.

\revised{Apart from the conventional approaches for specification generation, we further add the LLM-based approaches.}

\noindent \textbf{Few-shot LLM}. They refer to the large language model i.e., \verb|gpt-3.5| with few-shot settings in our work to generate specifications. Under the few-shot settings, the LLM is queried only once to obtain the final result. We set 0-shot, 2-shot, and 4-shot for few-shot comparison. 

\noindent \textbf{Conversational}. It refers to the generation technique described in Section~\ref{sec:conversation-driven_generation}. Conversational generation iteratively queries the LLM to refine its results, with error information provided to the LLM as feedback on each iteration. The conversational setting is based on 4-shot examples. Other settings remain the same with \tool.

\noindent \textbf{AutoSpec~\cite{wen2024enchanting}}. \revised{It is a recent technique for specification generation combining LLMs and static analysis. 
AutoSpec first decomposes the input program into its components, upon which a hierarchy graph is built. For each component, AutoSpec queries the LLM for corresponding specifications respectively. Eventually, specifications for all components are combined to obtain the overall result, which is presented to a verifier for correctness validation. The progress is repeated iteratively until the result is successfully verified.}

\vspace{-1mm}
\subsection{Evaluation Metrics}\label{sec:metrics}      
Following the previous works~\cite{alshnakat2020constraint, ghosal2023active}, we use the metric of \textit{Number of Passes} for assessment. We further add more metrics for comprehensive evaluation.


\noindent \textbf{Number of Passes}. It defines the number of programs for which the generated specifications of an approach pass the validation by the verifier. For a program, we consider the specifications that pass the verifier as the correct specifications. 


\noindent \textbf{Success Probability}. It is used to evaluate the model-based approaches. The randomness inherent in the content generated by large language models may introduce a certain level of contingency in a successful generation. Thus, we use the success probability for the measurement. For a test program, it is calculated by $\frac{N_{sucess}}{N_{attempt}}$ where $N_{sucess}$ denotes the number of successful generations of verifiable specifications and $N_{attemp}$ denotes a fixed number of trials in total (10 times in \tool).


\noindent \textbf{Number of Verifier Calls}. It is used to evaluate the efficiency of our approach. We propose a heuristic selection algorithm to prioritize the important candidates for the verifier to verify. To evaluate the effectiveness of the proposed selection algorithm, we use the number of verifier calls as the evaluation metric. 


\noindent \textbf{User Rating}. It aims to measure the semantic quality of the generated specifications. We invited 15 Ph.D. students who are experts in Java programming language to rate the specifications generated by different approaches. The research is conducted using the Likert Scale~\cite{nemoto2014likert}, where students are required to give a rating from one point to five points for each case according to a reference rating criteria.

\vspace{-1mm}
\section{Experimental Results}
\subsection{\lz{RQ1: Comparison with Baselines}}\label{sec:RQ1}

\begin{table*}[!t]
\centering
\caption{Number of programs that successfully pass the verifier and average success probability.}
\vspace{-2mm}
\label{tab:RQ1}
\scalebox{0.9}{
\begin{tabular}{cc|cc|cccccccccc|cc}
\midrule
\multicolumn{2}{c|}{\multirow{5}{*}{Approach}}                                                        & \multicolumn{2}{c|}{\multirow{2}{*}{\begin{tabular}[c]{@{}c@{}}\\SV-COMP\\ (265)\end{tabular}}} & \multicolumn{10}{c|}{\textit{SpecGenBench}}                                                                                                                                                                                                                                                                                                                                                                                            & \multicolumn{2}{c}{\multirow{2}{*}{\begin{tabular}[c]{@{}c@{}}\\Overall\\ (385)\end{tabular}}} \\ \cmidrule{5-14}
\multicolumn{2}{c|}{}                                                                                 & \multicolumn{2}{c|}{}                                                                         & \multicolumn{2}{c|}{\begin{tabular}[c]{@{}c@{}}Sequential\\ (26)\end{tabular}} & \multicolumn{2}{c|}{\begin{tabular}[c]{@{}c@{}}Branched\\ (23)\end{tabular}} & \multicolumn{2}{c|}{\begin{tabular}[c]{@{}c@{}}Single-path Loop\\ (24)\end{tabular}} & \multicolumn{2}{c|}{\begin{tabular}[c]{@{}c@{}}Multi-path Loop\\ (26)\end{tabular}} & \multicolumn{2}{c|}{\begin{tabular}[c]{@{}c@{}}Nested Loop \\ (21)\end{tabular}} & \multicolumn{2}{c}{}                                                                         \\
\multicolumn{2}{c|}{}                                                                                 & Num.                                        & Prob.                                           & Num.                      & \multicolumn{1}{c|}{Prob.}                         & Num.                     & \multicolumn{1}{c|}{Prob.}                        & Num.                         & \multicolumn{1}{c|}{Prob.}                            & Num.                        & \multicolumn{1}{c|}{Prob.}                            & Num.                                 & Prob.                                     & Num.                                        & Prob.                                          \\ \midrule
\multicolumn{2}{c|}{Daikon}                                                                           & 51                                          & -                                               & 10                        & \multicolumn{1}{c|}{-}                             & 10                       & \multicolumn{1}{c|}{-}                            & 0                            & \multicolumn{1}{c|}{-}                                & 1                           & \multicolumn{1}{c|}{-}                                & 0                                    & -                                         & 72                                          & -                                              \\ \midrule
\multicolumn{2}{c|}{Houdini}                                                                          & 56                                          & -                                               & 14                        & \multicolumn{1}{c|}{-}                             & 11                       & \multicolumn{1}{c|}{-}                            & 10                           & \multicolumn{1}{c|}{-}                                & 4                           & \multicolumn{1}{c|}{-}                                & 3                                    & -                                         & 98                                          & -                                              \\ \midrule
\multicolumn{1}{c|}{\multirow{3}{*}{\begin{tabular}[c]{@{}c@{}}Few-shot\\ LLM\end{tabular}}} & 0-shot & 81                                          & 18.28\%                                         & 23                        & \multicolumn{1}{c|}{74.19\%}                       & 17                       & \multicolumn{1}{c|}{58.55\%}                      & 5                            & \multicolumn{1}{c|}{7.08\%}                           & 7                           & \multicolumn{1}{c|}{18.13\%}                          & 2                                    & 3.33\%                                    & 135                                         & 22.93\%                                        \\
\multicolumn{1}{c|}{}                                                                        & 2-shot & 83                                          & 18.79\%                                         & 20                        & \multicolumn{1}{c|}{61.06\%}                       & 17                       & \multicolumn{1}{c|}{53.91\%}                      & 8                            & \multicolumn{1}{c|}{19.29\%}                          & 13                          & \multicolumn{1}{c|}{26.58\%}                          & 4                                    & 3.81\%                                    & 145                                         & 23.48\%                                        \\
\multicolumn{1}{c|}{}                                                                        & 4-shot & 94                                          & 19.40\%                                         & 23                        & \multicolumn{1}{c|}{73.85\%}                       & 20                       & \multicolumn{1}{c|}{57.33\%}                      & 10                           & \multicolumn{1}{c|}{23.40\%}                          & 12                          & \multicolumn{1}{c|}{24.95\%}                          & 5                                    & 6.24\%                                    & 164                                         & 25.25\%                                        \\ \midrule
\multicolumn{2}{c|}{Conversational}                                                                   & 146                                         & 30.95\%                                         & 23                        & \multicolumn{1}{c|}{82.49\%}                       & 20                       & \multicolumn{1}{c|}{75.43\%}                      & 12                           & \multicolumn{1}{c|}{27.02\%}                          & 13                          & \multicolumn{1}{c|}{35.38\%}                          & 4                                    & 9.20\%                                    & 218                                         & 35.95\%                                        \\ \midrule
\multicolumn{2}{c|}{\revised{AutoSpec}}                                                                         & \revised{156}                                         & \revised{\textbf{42.26\%}}                                         & \revised{24}                        & \multicolumn{1}{c|}{\revised{85.38\%}}                       & \revised{\textbf{21}}              & \multicolumn{1}{c|}{\revised{\textbf{85.20\%}}}                      & \revised{22}                           & \multicolumn{1}{c|}{\revised{57.00\%}}                          & \revised{16}                          & \multicolumn{1}{c|}{\revised{35.38\%}}                          & \revised{8}                                    & \revised{12.38\%}                                   & \revised{247}                                         & \revised{46.13\%}                                        \\ \midrule
\multicolumn{2}{c|}{\tool}                                                                             & \textbf{179}                                & 40.41\%                                & \textbf{24}               & \multicolumn{1}{c|}{\textbf{92.31\%}}              & 20                       & \multicolumn{1}{c|}{79.57\%}             & \textbf{23}                  & \multicolumn{1}{c|}{\textbf{73.75\%}}                 & \textbf{20}                 & \multicolumn{1}{c|}{\textbf{60.38\%}}                 & \textbf{13}                          & \textbf{36.55\%}                          & \textbf{279}                                & \textbf{59.97\%}                               \\ \midrule
\end{tabular}
}
\vspace{-5mm}
\end{table*}

The experimental results are presented in Table~\ref{tab:RQ1} where \textit{Num.} denotes the number of successfully handled programs, and \textit{Prob.} denotes the average success probability. Each LLM-based approach is granted 10 trials for each program.

\noindent \textbf{Performance on SV-COMP.} From Table~\ref{tab:RQ1}, we can find that on the SV-COMP dataset (265 programs in total), Houdini handled 56 programs, which is more than Daikon. However, both underperform LLM-based approaches even in LLM's simplest setting i.e. 0-shot setting, which handled 81 programs, demonstrating the feasibility of employing LLMs for formal program specification generation. Among the LLM-based approaches with few-shot settings, as the number of given few-shot examples increases, the number of programs that LLM can generate verifiable specifications is also increased, substantiating the effectiveness of the few-shot examples. Based on the 4-shots examples in the initial prompt, the multi-turn conversational manner (in Section~\ref{sec:conversation-driven_generation}) can further improve the performance, with 146 programs handled, compared to 94 programs of 4-shot LLM. \revised{Combining LLM-generated results and static analysis techniques, AutoSpec achieves enhanced results with 10 more programs handled.} Lastly, \tool outperforms all baseline approaches, with the number of programs that generated verifiable specifications increasing to 179.

\noindent \textbf{Performance on \textit{SpecGenBench}.} For deeper insights into the performance of different approaches on different types of programs, we further investigate the results on \textit{SpecGenBench}. Similar to SV-COMP, Houdini outperforms Daikon, especially in generating specifications for complex program structures such as loops, Houdini exhibits certain abilities. \revised{In terms of the LLM-based approaches, we find that \tool, AutoSpec, and the conversational approach are all competent in generating specifications for relatively simple programs such as sequential and branched. AutoSpec also demonstrates impressive ability on programs with Single-path Loops, benefiting from the code decomposition technique adopted.} However, when it comes to generating specifications for programs with more complex structures such as Multi-path Loops and Nested Loops, \tool has a clear advantage. The benefits are from our designed mutation-based specification generation (Section~\ref{sec:palusible_invariant_generation}), which can correct the erroneous output of the large language model to generate the verifiable specifications. Although \tool can generate more accurate specifications for different loop structures, it has a relatively poor performance in generating verifiable specifications \revised{for programs with nested loops}. 

For LLM-based approaches, we further use \revised{success probability} to evaluate the probability of successful generation for a program in 10 times trials due to the randomness inherent in the generated content by LLMs. \revised{We can observe that \tool achieves an overall success probability of 59.97\%, which is significantly higher than the values of conversational generation and AutoSpec (35.95\% and 46.13\% respectively). AutoSpec is not equipped with any feedback-refine mechanism. Although conversational generation can refine the intermediate results through conversation with LLM, there still exists detailed errors that cannot be fixed, since root causes of errors are difficult for LLMs to reason. Compared to these approaches, \tool, equipped with conversational generation and mutation-based generation, is more reliable with higher probabilities and lower chances of randomness to generate verifiable specifications.} From the results from SV-COMP and \textit{SpecGenBench}, we can find that our proposed approach is orthogonal to different datasets.

\begin{figure}[!t]
    \centering
    \includegraphics[width=0.28\textwidth]{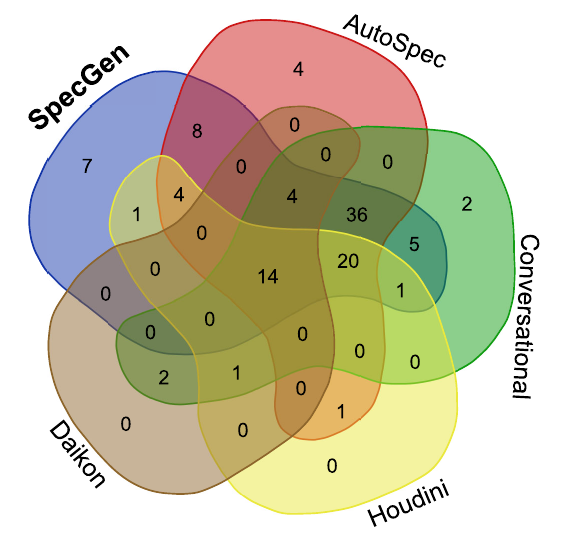}
    \vspace{-2mm}
    \caption{Venn diagram of verifiable programs.}
    \label{fig:venn}
    \vspace{-6mm}
\end{figure}

Presented in Fig.~\ref{fig:venn} is the Venn diagram of the programs in \textit{SpecGenBench} for which \tool and baseline approaches successfully generated verifiable specifications. It is noteworthy that \tool generates verifiable specifications for 7 programs that other baselines fail to yield, where 5 are from the Nested Loop, with the rest 2 from the Single-path Loop. Further investigation of these programs reveals that they are relatively complicated and challenging to handle.

\revised{Overall, it takes on average 7.07 verifier calls for the conversational approach per execution, whereas the figure is 16.19 and 15.51 for AutoSpec and \tool, respectively. Under our experimental settings, the maximum number of rounds for conversation is set to 10, so conversational generation naturally finishes execution faster than the other two approaches, but with relatively poor performance. In comparison, AutoSpec and \tool take more verifier calls to filter the LLM-generated results and produce more reliable results. Specifically, \tool displays a slight advantage over AutoSpec, indicating that \tool can achieve better performance compared to AutoSpec in the same or even shorter period of time.}

\vspace{-1mm}
\begin{tcolorbox}
\vspace{-1mm}
\textbf{RQ1:} \tool outperforms all baselines on two different datasets, generating verifiable specifications for 279 out of 385 programs with the highest success probability among all LLM-based approaches. In comparison, the number of programs handled by Daikon, Houdini, Conversational generation, and AutoSpec is 72, 98, 218, and 247, respectively.
\vspace{-1mm}
\end{tcolorbox}

\begin{table}[!t]
\centering
\caption{Effectiveness of different types of mutations.}
\vspace{-1mm}
\scalebox{0.68}{
\begin{tabular}{c|ccccc|c|c}
\toprule
\multirow{3}{*}{\begin{tabular}[c]{@{}c@{}}\\Approach\end{tabular}}                                  & \multicolumn{5}{c|}{\textit{SpecGenBench}}                                                                                                                                                                                                                                                                                         & \multirow{3}{*}{\begin{tabular}[c]{@{}c@{}}\\SV-COMP\\ (265)\end{tabular}} & \multirow{3}{*}{\begin{tabular}[c]{@{}c@{}}\\Total\\ (385)\end{tabular}} \\ \cmidrule{2-6}
                                                          & \begin{tabular}[c]{@{}c@{}}Sequential\\ (26)\end{tabular} & \begin{tabular}[c]{@{}c@{}}Branched\\ (23)\end{tabular} & \begin{tabular}[c]{@{}c@{}}Single-path\\ (24)\end{tabular} & \begin{tabular}[c]{@{}c@{}}Multi-path\\ (26)\end{tabular} & \begin{tabular}[c]{@{}c@{}}Nested\\ (21)\end{tabular} &                                                                            &                                                                        \\ \midrule
\begin{tabular}[c]{@{}c@{}}w/o Predicative\end{tabular} & 24                                                        & 20                                                      & 20                                                                & 19                                                               & 9                                                            & 167                                                                        & 259                                                                    \\
\begin{tabular}[c]{@{}c@{}}w/o Logical\end{tabular}     & 24                                                        & 18                                                      & 14                                                                & 18                                                               & 10                                                           & 151                                                                        & 235                                                                    \\
\begin{tabular}[c]{@{}c@{}}w/o Comparative\end{tabular} & 24                                                        & 19                                                      & 13                                                                & 12                                                               & 7                                                            & 148                                                                        & 223                                                                    \\ 
\begin{tabular}[c]{@{}c@{}}w/o Arithmetic\end{tabular}  & 23                                                        & 19                                                      & 18                                                                & \textbf{21}                                                      & 11                                                           & 170                                                                        & 262                                                                    \\ \midrule
\tool                                      & \textbf{24}                                               & \textbf{20}                                             & \textbf{23}                                                       & 20                                                               & \textbf{13}                                                  & \textbf{179}                                                               & \textbf{279}                                                           \\ \bottomrule
\end{tabular}
}
\label{tab:ablation}
\vspace{-5mm}
\end{table}

\subsection{\lz{RQ2: Ablation Study on Mutation Types}} \label{sec:ablation}
We conduct an ablation study to evaluate the effectiveness of different mutation types in \tool. The results are shown in Table~\ref{tab:ablation} where w/o \{*\} denotes the disabled mutation type.

We can find that \tool successfully generates verifiable specifications for 279 out of a total of 385 programs. \tool w/o Comparative addresses the least number of programs i.e., 223, indicating that the comparative mutation is the most important in the defined mutation operations. The main reason is the frequent usage of numerical variables in programs and the recurring need to bound their range in the specifications. 
\tool w/o Logical has the second least number of programs i.e., 235, indicating that the logical operators are also important to generate verifiable specifications. This is due to the necessity of combining two or more expressions for different properties with logical operations when specifying complex behaviors.
\tool w/o Predicative and \tool w/o Arithmetic have the most number of programs (259 and 262 respectively), which means both of them are less important than the comparative and logical mutation. We still consider them as they are applicable in certain situations. In some complex programs, specifications generated by the LLM are prone to have predicate errors. Hence, the predicative mutation will be useful. Similar cases exist when there are complicated numerical constraints on variables, where mutations on arithmetic operators turn out to be helpful.

Further analyzing the effectiveness of the mutation type for different kinds of programs, we can find that \tool w/o Comparative handles an especially lower number of programs in the loop category including single-path, multi-path, and nested loop. It is due to the rigid demand of scope bounding for loop variables when loops are involved. Scope bounding for loop variables is an intricate work where LLMs frequently make mistakes, substantiating the importance of comparative mutations. The performance of \tool w/o Predicative also drops on programs with nested loop, because of the relatively higher quantity and complexity of \verb|\forall| and \verb|\exists| statements involved in these nested programs.

\vspace{-1mm}
\begin{tcolorbox}
\vspace{-1mm}
\textbf{RQ2:} Each type of mutations contributes differently to \tool. The comparative mutation contributes the most to the performance while the predicative and arithmetic are less important. When combining them together, \tool achieves the best performance.
\vspace{-1mm}
\end{tcolorbox}
\vspace{-1mm}

\begin{table}[!t]
\centering
\caption{Average numbers of verifier calls in a single run under different specification selection strategies in Section~\ref{sec:spec-selection}.}
\vspace{-1mm}
\scalebox{0.72}{
\begin{tabular}{c|c|ccccc|c}
\toprule
\multirow{3}{*}{Strategy} & \multirow{3}{*}{SV-COMP} & \multicolumn{5}{c|}{\textit{SpecGenBench}}                                                                                                                            & \multirow{3}{*}{Total} \\ \cmidrule{3-7}
                          &                          & Sequential & Branched & \begin{tabular}[c]{@{}c@{}}Single-path\end{tabular} & \begin{tabular}[c]{@{}c@{}}Multi-path\end{tabular} & Nested &                          \\ \midrule
Random                    & 9.41                     & 2.70       & 2.32     & 32.62                                                      & 55.16                                                     & 41.97       & 18.44                    \\
Heuristic                 & \textbf{8.91}                     & \textbf{2.59}       & \textbf{2.16}     & \textbf{24.21}                                                      & \textbf{43.58}                                                     & \textbf{34.99}       & \textbf{15.51}                    \\ \bottomrule
\end{tabular}
}
\label{tab:efficiency}
\vspace{-4mm}
\end{table}

\vspace{-1mm}
\subsection{RQ3: Effectiveness of Selection Strategy}
In Section~\ref{sec:spec-selection}, we design the heuristic selection strategy to improve the efficiency of verification. We also conduct an experiment to compare with the random selection strategy. Specifically, when a candidate specification is refuted by the verifier, we randomly select another specification from $E_f$ for replacement. To compare with different strategies, for each program that successfully generates verifiable specifications by \tool, we run \tool 5 times to obtain the average number of verifier calls as the evaluation metric.

\noindent \textbf{Performance on SV-COMP.} Using the random selection strategy makes 9.41 verifier calls on average while the heuristic selection strategy takes 8.91 calls, resulting in an improvement of 5.30\%. The improvement is relatively modest and the values denote that only less than 10 verifier calls on average are used to generate verifiable specifications for programs in SV-COMP. The main reason is that the number of rounds for conversation is set to 10 in \tool, the specifications for these programs tend to be successfully generated within the conversation module (Section~\ref{sec:conversation-driven_generation}). It is before the selection strategy used in the mutation-based specification generation (Section~\ref{sec:palusible_invariant_generation}) comes into effect.

\noindent \textbf{Performance on \textit{SpecGenBench}.} Using the random selection strategy takes 36.20 verifier calls on average while the heuristic selection strategy takes 28.51 calls in terms of five categories in \textit{SpecGenBench}. Hence, the heuristic selection strategy achieves an improvement of 21.23\%. Furthermore, we can observe that the improvements in different categories of programs vary significantly. The improvements in the loop categories including single-path, multi-path, and nested loop are more significant than sequential and branched. The main reason is that generating specifications for loop-containing programs is challenging, and usually requires more iterations to obtain verifiable specifications. In this case, a good selection strategy often highlights advantages more effectively. However, the improvements in sequential and branched categories are fewer. \lz{The reason is similar to SV-COMP, where the high efficiency of conversational generation on sequential and branched programs makes the selection strategies invalid.} Nevertheless, loop structures are common in programs, thus a heuristic selection strategy to improve the validation efficiency is still helpful and necessary.

\begin{tcolorbox}
\vspace{-1mm}
\textbf{RQ3:} The heuristic selection strategy effectively improves the efficiency of \tool. It is especially useful when generating specifications for programs with more complex structures such as loops.
\vspace{-1mm}
\end{tcolorbox}

\vspace{-1mm}
\subsection{RQ4: User Study on the Quality of Specifications}
A user study is conducted to evaluate the semantic quality of the generated specifications. 15 Ph.D. students are invited to rate the specifications generated by different approaches. A detailed description of the rating process is given in Section~\ref{sec:metrics}. We selected the 15 programs from the dataset of \textit{SpecGenBench} that can be handled by all of Houdini, Daikon, and \tool. Apart from the specifications generated by these approaches, we also add the ground truth as a reference. \revised{The specifications are kept anonymous to the students, disclosing no information about the sources of the specifications.} The rating scores are presented in Table~\ref{tab:user_rating}, \sql{where the score for full marks is 5.}


We can observe that the ground truth specifications (oracle) receive an average rating score of 4.83, indicating that the semantics of these programs can be described comprehensively through JML specifications. Furthermore, the specifications generated by \tool received a rating score of 4.54, which is close to the oracle, indicating that the generated specifications by \tool can also describe the real behaviors of the input program more fully. Among the 15 programs, all rating scores given to \tool are above 3, with the lowest rating being 3.33, meaning that in the worst case, \tool can still generate non-trivial specifications about the properties of the input program. In comparison, the specifications generated by Houdini and Daikon received an average rating score of 2.32, reflecting the semantic weakness in these specifications. \revised{Houdini and Daikon rely on pre-defined templates, which are in fact independent from the input program and can only cover a limited number of specification patterns. Consequently, the specifications produced are often simplistic and trivial, involving only a narrow range of variables and operators, making it difficult to capture the actual behavior and functionality of the input program precisely. Unlike traditional approaches that rely on a fixed set of templates, \tool utilizes the code comprehension capabilities of LLMs, which can cover a larger range of scenarios and generate targeted specifications that more closely match the semantics of the input program.}


\vspace{-1mm}
\begin{tcolorbox}
\vspace{-1mm}
\textbf{RQ4:} \tool received an average rating score of 4.54, which is close to the 4.83 of the oracle specifications, demonstrating the ability to accurately characterize the real program behaviors and generate specifications with comprehensive program semantics.
\vspace{-1mm}
\end{tcolorbox}
\vspace{-2mm}

\begin{table}[!t]
\centering
\caption{Average rating scores on the generated specifications by different approaches.}
\vspace{-1mm}
\scalebox{0.8}{
\begin{tabular}{c|cccc}
\toprule
Test case          & Houdini & Daikon & SpecGen       & Oracle \\ \midrule
Absolute           & 3.50    & 3.36   & \textbf{4.85} & 5.00   \\
AddLoop            & 2.40    & 1.33   & \textbf{4.57} & 5.00   \\
Conjunction        & 4.50    & 3.50   & \textbf{5.00} & 5.00   \\
ConvertTemperature & 2.33    & 2.50   & \textbf{5.00} & 5.00   \\
Disjunction        & 2.50    & 3.50   & \textbf{5.00} & 5.00   \\
FizzBuzz           & 2.63    & 2.86   & \textbf{5.00} & 5.00   \\
IsCommonFactor     & 2.00    & 4.13   & \textbf{4.14} & 4.71   \\
IsPalindrome       & 1.83    & 1.17   & \textbf{4.75} & 5.00   \\
IsSubsequence      & 2.43    & 1.13   & \textbf{4.14} & 4.00   \\
IsSuffix           & 2.20    & 1.50   & \textbf{4.33} & 4.63   \\
MulLoop            & 1.88    & 1.25   & \textbf{3.33} & 5.00   \\
MySqrt             & 2.00    & 2.80   & \textbf{3.75} & 4.25   \\
Perimeter          & 1.00    & 2.80   & \textbf{4.78} & 5.00   \\
SmallestEvenMul    & 2.57    & 1.00   & \textbf{4.50} & 5.00   \\
Swap               & 1.00    & 2.00   & \textbf{5.00} & 4.88   \\ \midrule
Average            & 2.32    & 2.32   & \textbf{4.54} & 4.83   \\ \bottomrule
\end{tabular}
}
\label{tab:user_rating}
\vspace{-5mm}
\end{table}

\section{Discussion}
\subsection{\revised{Performance on Real-world Programs}}

\begin{table*}[!htb]
\centering
\caption{Performance on programs collected from 9 repositories in Defects4J.}
\vspace{-2mm}
\label{tab:defects4j}
\scalebox{0.69}{
\begin{tabular}{c|cccccccccccccccccc|cc}
\toprule
\multirow{2}{*}{Approaches} & \multicolumn{2}{c}{\begin{tabular}[c]{@{}c@{}}chart\\ (7)\end{tabular}} & \multicolumn{2}{c}{\begin{tabular}[c]{@{}c@{}}cli\\ (5)\end{tabular}} & \multicolumn{2}{c}{\begin{tabular}[c]{@{}c@{}}codec\\ (4)\end{tabular}} & \multicolumn{2}{c}{\begin{tabular}[c]{@{}c@{}}compress\\ (6)\end{tabular}} & \multicolumn{2}{c}{\begin{tabular}[c]{@{}c@{}}jackson\\ (7)\end{tabular}} & \multicolumn{2}{c}{\begin{tabular}[c]{@{}c@{}}jxpath\\ (6)\end{tabular}} & \multicolumn{2}{c}{\begin{tabular}[c]{@{}c@{}}lang\\ (7)\end{tabular}} & \multicolumn{2}{c}{\begin{tabular}[c]{@{}c@{}}math\\ (4)\end{tabular}} & \multicolumn{2}{c|}{\begin{tabular}[c]{@{}c@{}}time\\ (4)\end{tabular}} & \multicolumn{2}{c}{\begin{tabular}[c]{@{}c@{}}Total\\ (50)\end{tabular}} \\
                       & Num.                             & Prob.                                & Num.                            & Prob.                               & Num.                             & Prob.                                & Num.                              & Prob.                                  & Num.                              & Prob.                                 & Num.                             & Prob.                                 & Num.                            & Prob.                                & Num.                            & Prob.                                & Num.                             & Prob.                                & Num.                              & Prob.                                \\ \midrule
Daikon                 & 3                                & -                                    & 3                               & -                                   & 0                                & -                                    & 1                                 & -                                      & 3                                 & -                                     & 2                                & -                                     & 2                               & -                                    & 1                               & -                                    & 0                                & -                                    & 15                                & -                                    \\ \midrule
4-shot LLM             & 1                                & 7.14\%                                & 2                               & 7.33\%                               & 2                                & 9.71\%                                & 3                                 & 11.67\%                                 & 3                                 & 9.52\%                                 & 1                                & 2.78\%                                 & 4                               & 11.67\%                               & 3                               & 17.50\%                               & 1                                & 5.00\%                                & 20                                & 8.97\%                                \\ \midrule
Conversational         & 4                                & 47.62\%                               & 4                               & 60.00\%                              & 3                                & 41.67\%                               & 1                                 & 11.11\%                                 & 2                                 & 19.05\%                                & 5                                & 55.56\%                                & 3                               & 28.57\%                               & 3                               & \textbf{66.67\%}                               & 3                                & \textbf{41.67\%}                               & 28                                & 39.33\%                               \\ \midrule
\textit{SpecGen}       & \textbf{6}                       & \textbf{68.57\%}                      & \textbf{4}                      & \textbf{68.00\%}                     & \textbf{4}                       & \textbf{65.00\%}                      & \textbf{4}                        & \textbf{36.67\%}                        & \textbf{4}                        & \textbf{42.86\%}                       & \textbf{5}                       & \textbf{63.33\%}                       & \textbf{5}                      & \textbf{65.71\%}                      & \textbf{3}                      & 45.00\%                      & \textbf{3}                       & 35.00\%                      & \textbf{38}                       & \textbf{55.20\%}                      \\ \bottomrule
\end{tabular}
}
\vspace{-5mm}
\end{table*}

\revised{
To further evaluate the performance of \tool on real-world programs, we collect programs involved in Defects4J~\cite{just2014defects4j}, a well-known dataset of reproducible bugs within open-source repositories. During the collection process, we only consider individual files with no dependency on third-party libraries or other files in the repository. This ensures that all the collected files can be properly executed and verified outside the repository. Eventually, 50 Java source files from 9 repositories are collected. The average line of code and cyclomatic complexity of the collected programs are 374.78 and 18.29, respectively. We follow the same experimental settings in Section~\ref{sec:implementation} for experiments. Note that we only aim to evaluate the verifiability of the generated specifications in the same way Section~\ref{sec:RQ1} does, so the ground truth specifications of the programs are not prepared.
}

\revised{Table~\ref{tab:defects4j} shows the performance of \tool and other baseline methods on the programs extracted from Defects4J. Although Daikon underperforms the LLM-based approaches, it still exhibits certain abilities in processing real-world programs. This is due to the existence of some simplistic methods within real-world class definitions, such as those retrieving the value of a certain class member without doing anything else, which Daikon is capable of handling. Compared to Daikon, the LLM-based approach with the simplest setting, i.e. 4-shot, achieved 5 more programs handled. Based on the few-shot learning technique, the conversational approach further achieved 28 programs handled. Lastly, \tool succeeds in handling 38 out of the 50 programs, with an average success probability of 55.20\%, displaying decent capabilities in handling real-world programs.
}

\vspace{-1mm}
\subsection{Threats to Validity}
\noindent \textbf{Internal Validity.}
First, the prompts we used to communicate with the LLM may affect our results. To mitigate it, we refer to Xia et al.~\cite{xia2023conversational} to design the prompt. We plan to investigate the effect of different prompts in the future. Second, a potential threat lies in the risk of data leakage. Our constructed dataset \textit{SpecGenBench} consists of 100 programs with expert-written specifications and 20 programs with their corresponding specifications from Nilizadeh et al.~\cite{Nilizadeh2021AprFormalMethods}. The former does not have the issue of data leakage as the specifications are written by experts in our research. However, since \verb|gpt-3.5| does not release its model as well as the training data, the latter 20 programs from the existing dataset may have the risk. The used dataset SV-COMP also has this risk. Nevertheless, through our observation of \tool on these programs, we have never spotted a situation where the output of \tool is the same as the existing oracle. Hence, we believe this threat is limited. Furthermore, even if we remove these potentially risky programs, \tool still successfully handles 87 programs in the remaining 100 programs, which is also the best. 

\noindent \textbf{External Validity.}
One of the external threats lies in the accuracy of the verifier (OpenJML). Due to the implementation flaws in OpenJML, there may be cases where some correct specifications fail to pass verification. This is an inevitable problem that other verifiers~\cite {ahrendt2005key} have to face as well. The reason lies in the undecidability of automatic software verification~\cite{abdulla1996undecidable, mathur2020s}. Even in such a situation, \tool achieves impressive performance with the majority of testcases successfully generated verifiable specifications. Another threat is the potential bias of the hand-written specifications by experts. To mitigate this, we follow the procedure in Nilizadeh et al.~\cite{Nilizadeh2021AprFormalMethods}. First, the selected experts should have rich experience in writing specifications. Second, the initially chosen specifications should be verifiable by the verifier. Last, if multiple experts have written the specifications that pass the verifier, another expert is responsible for selecting one. 


\section{Related Work}

\noindent \textbf{Large Language Models.} With the advancement of generative AI, Large Language Models (LLMs) have emerged as a formidable force and have quickly found widespread applications. LLMs are characterized by their immense parameter scale and training dataset size~\cite{zhao2023survey}.
An important feature of LLMs is their ability for in-context learning~\cite{brown2020language}, which enhances the coherence between the context and the output of LLMs. The learning ability gives rise to a unique usage of LLMs known as prompting~\cite{liu2023pre}, where a natural language description of the intended downstream task is provided to the LLM before assigning it the task. LLMs initially demonstrated remarkable capabilities in the field of Natural Language Processing (NLP)~\cite{min2023recent}, excelling in tasks such as document classification~\cite{hegselmann2023tabllm}, text summarization~\cite{yang2023exploring}, and machine translation~\cite{zhang2023prompting}. They are also widely deployed in various software engineering tasks~\cite{ma2023scope,hou2023large}, including software testing~\cite{deng2023large,xia2023universal}, code generation~\cite{zeng2022extensive,liu2023your} and code summarization~\cite{ahmed2022few}. Compared with these works, our goal is to employ LLMs for the automated generation of program specifications, which is important in formal methods. 

\noindent \textbf{Program Specification Generation.} The research on program specification generation can be categorized into two types: natural language specification generation and formal specification generation.
Natural language specification generation primarily manifests as code summarization~\cite{zhu2019automatic}, a process of automatically generating accurate, human-readable descriptions of code functionality. Numerous efforts have been made to utilize machine learning methods for code summarization~\cite{iyer2016summarizing,ahmad2020transformer, liu2020retrieval,fernandes2018structured}. Formal specification generation primarily takes the form of the generation of program invariants, the formal language representations of properties that a program is guaranteed to satisfy at a certain program point. In invariant generation, a large amount of research focuses on the generation of loop invariants~\cite{garg2016learning,ryan2019cln2inv,si2018learning,janssen2023can,chakraborty2023ranking,kamath2023finding,hellendoorn2019my}, while the rest of the works attempt to generate invariants of other forms, e.g. pre-conditions~\cite{cousot2013automatic,ghosal2023active}, post-conditions~\cite{moy2010modular,wei2011inferring,alshnakat2020constraint,molina2021evospex,molina2022fuzzing}, assertion-based invariants~\cite{terragni2020evolutionary} and finite automata~\cite{christodorescu2007mining,aarts2012automata,murali2017bayesian}. \revised{With the development of Large Language Models, there have also been efforts employing LLMs to generate program specifications. Wen et al.~\cite{wen2024enchanting} combine LLMs with static analysis techniques, including code decomposition, to generate verifiable program specifications. Pei et al.~\cite{pei2023can} utilize fine-tuning to enhance the performance of LLMs on specification generation tasks. Concerning the difficulty of selecting correct specifications from the massive LLM-generated results, Chakraborty et al.~\cite{chakraborty2023ranking} propose a ranking algorithm that can distinguish correct inductive invariants from incorrect attempts based on the problem definition.} Among these works, artifacts of Ghosal et al.~\cite{ghosal2023active} and Pei et al.~\cite{pei2023can} are not publicly available, the artifact of Alshnakat et al.~\cite{alshnakat2020constraint} is built for C code and Frama-C contracts, and the grammar for specifications of Molina et al.~\cite{molina2021evospex,molina2022fuzzing} involves specific features, which cannot be trivially translated into equivalent JML, thus we cannot include them for comparison. Compared to these works, \tool utilizes the code comprehension capability of LLMs for program specification generation in a conversational manner, further followed by the mutation-based approach for enhancement.


\vspace{-1mm}
\section{Conclusion}
In this paper, we introduced \tool, a novel approach that utilizes the Large Language Model for formal program specification generation. Leveraging the code comprehension ability of LLMs as well as the well-designed mutation-based specification generation component, our approach is capable of accurately capturing the behaviour and functionality of input programs to generate accurate specifications. A comprehensive evaluation between \tool and other baselines is conducted on two different datasets, the benchmark for the Java category of SV-COMP, and a more diverse and manually constructed dataset containing 120 programs. The extensive experimental results haver demonstrated that our approach significantly outperforms the baseline approaches, with the ability to effectively articulate program behaviors.


\vspace{-1mm}
\section*{Acknowledgment}
We are grateful for the constructive feedback of all the anonymous reviewers to improve this manuscript. The authors from Nanjing University are supported in part by the Leading-edge Technology Program of Jiangsu Natural Science Foundation (No. BK20202001), the National Natural Science Foundation of China (No. 62232008, 62172200), and the Postgraduate Research \& Practice Innovation Program of Jiangsu Province (No. KYCX24\_0237).

\bibliographystyle{IEEEtran}
\bibliography{ref}

\end{document}